\documentclass[aps,prl,twocolumn,groupedaddress,floatfix,showpacs]{revtex4}

\usepackage{subfigure}
\usepackage{color}
\usepackage{graphicx}
\usepackage{amssymb}
\usepackage{amsmath}
\usepackage{natbib}

\begin{document}

\newcommand{\eps}{\varepsilon}
\newcommand{\ud}{\mathrm{d}}
\newcommand{\pd}{\partial}
\newcommand{\vd}{\delta}
\newcommand{\Tr}{\mathrm{Tr}}
\newcommand{\T}{\mathrm{T}}
\newcommand{\bq}{\begin{equation}}
\newcommand{\eq}{\end{equation}}
\newcommand{\bea}{\begin{eqnarray}}
\newcommand{\eea}{\end{eqnarray}}
\newcommand{\bef}{\begin{figure}}
\newcommand{\ef}{\end{figure}}
\newcommand{\bdm}{\begin{displaymath}}
\newcommand{\edm}{\end{displaymath}}
\newcommand{\ba}{\begin{array}}
\newcommand{\ea}{\end{array}}

\newcommand{\Gph}{\Gamma^{\mathrm{ph}}}
\newcommand{\kk}{\mathbf{k}}
\newcommand{\QQ}{\mathbf{Q}}
\newcommand{\qq}{\mathbf{q}}
\newcommand{\BB}{\mathbf{B}}
\newcommand{\up}{\uparrow}
\newcommand{\dn}{\downarrow}

\newcommand{\uu}{\up\!\up}
\newcommand{\updn}{\up\!\dn}
\newcommand{\dnup}{\dn\!\up}
\newcommand{\dd}{\dn\!\dn}
\newcommand{\upd}{\up\dn}

\newcommand{\Tuu}{T_{\uu}}
\newcommand{\Tdd}{T_{\dd}}
\newcommand{\Tud}{T_{\updn}}
\newcommand{\Tdu}{T_{\dnup}}

\newcommand{\Gud}{G_{\up,\dn}}
\newcommand{\g}{\gamma}
\newcommand{\D}{\Delta}
\newcommand{\la}{\langle}
\newcommand{\ra}{\rangle}

\newcommand{\al}{\alpha}
\newcommand{\tz}{\theta_0}
\newcommand{\tR}{\theta_R}
\newcommand{\eso}{\eps_\text{so}}
\newcommand{\Eso}{E_\text{so}}
\newcommand{\Ef}{E_F}
\newcommand{\kf}{k_F}
\newcommand{\vf}{v_F}
\newcommand{\kfup}{k_{F\up}}
\newcommand{\kfdn}{k_{F\dn}}
\newcommand{\vfup}{v_{F\up}}
\newcommand{\vfdn}{v_{F\dn}}
\newcommand{\kfupdn}{k_{F_{\up,\dn}}}
\newcommand{\vfupdn}{v_{F_{\up,\dn}}}

\newcommand{\ddx}{\frac{\partial}{\partial x}}
\newcommand{\ddy}{\frac{\partial}{\partial y}}
\newcommand{\hs}{\hat\sigma}
\newcommand{\hp}{\hat p}
\newcommand{\kr}{k_R}
\newcommand{\ups}{|\!\!\up\ra}
\newcommand{\dns}{|\!\!\dn\ra}
\newcommand{\plus}{|+\ra}
\newcommand{\minus}{|-\ra}
\newcommand{\psim}{|\psi_-\ra}
\newcommand{\psip}{|\psi_+\ra}
\newcommand{\rights}{|\!\to\ra}
\newcommand{\lefts}{|\!\gets\ra}

\parskip=0pt

\title{Magnetically induced oscillations of the spin polarization in the Datta--Das geometry}
{\rm }


\author{Andriy H. Nevidomskyy}

\email[\vspace{-2mm}E-mail: ]{nevidomskyy@cantab.net}


\author{Karyn Le Hur}

\affiliation{D\'epartement de Physique, Universit\'e de Sherbrooke,
  Sherbrooke, Qu\'ebec, J1K 2R1, Canada}


\date{\today}

\begin{abstract}
The control of intrinsic magnetic degrees of freedom is very important
as it offers  a practical means to manipulate  and probe electron spin
transport. Tunable spin-orbit effect in quantum wires can in principle
serve as a means to achieve this goal.  Here, we investigate
within the scattering matrix approach the effect
of an in-plane magnetic field on the conductance of a quantum wire
in the Datta-Das geometry and
show  that  the  interplay  of  the spin-orbit  interaction  with the
magnetic  field  provides  enhanced  control over  the  electron  spin
polarization.
In  particular,  we predict  a  novel  effect of  magnetically induced
oscillations  of the  electron  spin  in a  certain  range of 
magnetic field.
\end{abstract}

\pacs{
 71.70.Ej,   
 72.25.-b, 
  72.25.Dc, 
  73.63.Nm  
\vspace{-2mm}
}


\maketitle

\vspace{-5mm}

Spin effects in transport embody a new branch in mesoscopic physics and
semiconductor spintronics~\cite{spintronics} with several
technological applications such as information storage,
magnetic sensors, and potentially quantum computation~\cite{qc_review00}.
One of the first devices that would control the electron
spins was proposed by Datta and Das~\cite{Datta-Das} who suggested the
use of the gate-controlled Rashba spin-orbit interaction~\cite{Rashba} as a means of varying
the spin polarization in one-dimensional quantum wires. The use of the
Rashba interaction in beam-splitter devices and its effect on the
shot noise have also been envisioned in the context of quantum entanglement~\cite{Egues02}.
%
%
The combined effect of the Zeeman field and spin-orbit coupling has
been the subject of a number of recent works~\cite{Serra05,Devillard05,Cahay03}, 
however they have concentrated on the band structure
issues~\cite{Serra05} or on
the spectrum of collective excitations~\cite{Devillard05}. The
resonances in transmission in the presence of magnetic field (and also disorder) have been
studied~\cite{Cahay03}, however the field dependence of the
conductance was not calculated. 
The latter dependence will be the main goal of this study.
%

Here we  consider the Datta-Das setup~\cite{Datta-Das} that consists
of a quantum wire with the Rashba~\cite{Rashba} spin-orbit interaction
(region II  in Fig.~\ref{Fig.Datta-Das}) coupled  to the ferromagnetic
leads~(regions  I  and III),  so that  a
spin-polarized  current  can be  injected  from  the  left and drained
on the right.  
The whole setup, including the leads, is then put into an in-plane magnetic field,
allowing us to avoid orbital as well as the Aharonov--Bohm effects~\cite{AB,ALG} that 
would have occured if the field were perpendicular to the device plane.
Applying the scattering matrix formalism, we demonstrate that the
interplay of the magnetic field with the spin-orbit effect gives rise to 
the novel spin oscillations that can be exploited to enhance
the control over the electron spin-polarization in the Datta-Das geometry.




The spin-orbit coupling is a relativistic effect \cite{Landau} 
resulting in the Hamiltonian $\hat H_{SO}\propto
\nabla V\cdot(\mathbf{\hat\sigma}\times\mathbf{\hat p})$, 
where $\mathbf{\hat p}$ is the electron momentum,
$\mathbf{\hat\sigma}$ are the Pauli matrices and V is the
electrostatic potential.
It has been shown that in most two-dimensional
heterostructures~\cite{Das90,Chen93} on which the Datta-Das device is
based, the Dresselhaus spin-orbit effect due to the microscopic crystal
field~\cite{Dresselhaus} is much smaller compared to the Rashba
effect caused by  the quantum-well asymmetry. 
Therefore we shall limit our discussion to the Rashba
Hamiltonian~\cite{Rashba}:
%
\bq
\hat H_R=\al(\mathbf{\hat\sigma}\times\mathbf{\hat p})_z/\hbar
= i\al\left(\hat\sigma_y\partial_x-\hat\sigma_x\partial_y \right).
\label{HR}
\vspace{-1mm}
\eq

The constant $\al$ absorbs in itself the electric field normal to the
semiconductor interface and takes value in the range
$(1-10)\times10^{-10}$~eV~cm for a large variety of systems
(InAs/GaSb \cite{Luo90},
In$_x$Ga$_{1-x}$As/In$_x$Al$_{1-x}$As \cite{Das90,Nitta} and
GaAs/Al$_x$Ga$_{1-x}$As \cite{Hassenkam97}) depending on the shape of the
confining quantum well.
The energy scale of the Rashba spin-orbit interaction is given by
$\Eso=\hbar^2 k_R^2/2m^*=\alpha^2 m^*/2\hbar^2$ and is of the order of
$10^{-4}-10^{-2}$~meV,
where $k_R=\al m^*/\hbar^2$ is the characteristic Rashba wave-vector.

\bef [!tbp]
\begin{center}{\includegraphics[width=8.6cm]{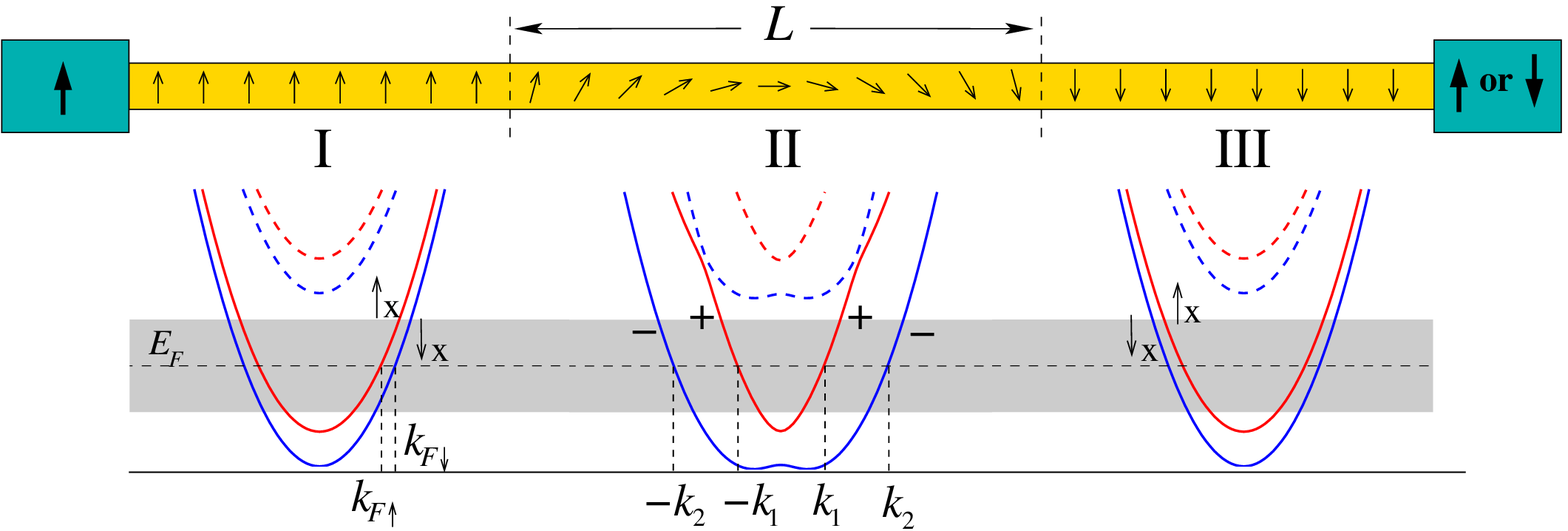}}
\end{center}
\vspace{-8mm}
\caption
{Datta-Das geometry and the sketch of the energy band
  structure in the applied magnetic field $\BB\|x$. The Rashba-active region of length $L$ 
  comprises the middle part of the wire.
  In the energy dispersion plots, two transverse bands (solid and
  broken lines) are shown, however the
  discussion is limited to the position of the Fermi
  level, $\Ef$, in the grey region, where only the lowest transverse channel
  is active. 
 The up and down arrows refer to the direction of $\BB$.
}
\label{Fig.Datta-Das}
\vspace{-5mm}
\ef

Let us consider the case of zero magnetic field first.
For electrons moving along the wire in the region~II
in Fig.~\ref{Fig.Datta-Das} we have  $p_y$=$\,0$ and the Rashba term  
$H_R=\alpha\hs_y \hp_x/\hbar$. Hence the electrons with opposite projections of
the $\hs_y$ component of the spin 
correspond to the different Rashba-split dispersion branches with
different Fermi wave-vectors $k_1$ and $k_2$. They will thus pick 
up different phases in 
the Rashba region of length $L$, and the transmission probability will
be governed by the phase difference $(k_2-k_1)L$. It is equal to 
$2\theta_R\equiv2k_RL=2\al Lm^*/ \hbar^2$ and is proportional to the
spin-orbit strength $\alpha$. 
For the equal
spin-polarizations of the source and drain, say, the
transmission of the wire reads~\cite{Datta-Das}: 
$\Tuu=\frac{1}{2}(1+\cos2\theta_R)$. 


\emph{Model}. In this work we shall consider the effect that the in-plane
magnetic field $B$ has on the transmission of the Datta-Das
setup. 
We start from the Hamiltonian 
%
%
%
\bq
\hat H = \frac{\hp^2}{2m^*} - \frac{\alpha}{\hbar}\hs_y\hp_x +
\D\, \hs_x +  V_c(y) + \frac{\alpha}{\hbar}\hs_x\hp_y,
\label{H}
\eq
where the Zeeman energy $\D=g\mu_B B/2$, and
$V_c(y)$ is the lateral confining potential.
Note that the field $\BB$ could also be applied along the
$z$ axis and would not modify the eigen-energies (when neglecting orbital
effects).

%

For narrow enough  wires the electron momentum is  aligned mostly with
the direction of the wire, so that $\la p_x\ra \gg \la p_y\ra$, and we
shall neglect the last spin-orbit term in the Hamiltonian (\ref{H}).
This results in a simple
picture  shown schematically in
Fig.~\ref{Fig.Datta-Das} of a number of non-interacting transverse
subbands with energies $E_n^y$ that 
can be obtained by diagonalizing the $y$-dependant part of the Hamiltonian.
It has been demonstrated~\cite{Moroz99,Governale04} however that this
approximation breaks down in the vicinity of the crossing of
the transverse subbands, since the spin-orbit interaction would mix
different transverse channels and lift the subband degeneracy, leading
to the band anticrossing.
Egues, Burkard and Loss~\cite{Egues03} have shown that
this anticrossing can be exploited in the Datta-Das setup to enhance
the spin control. Here we shall consider the realm of only the lowest 
subband, which can be achieved by tuning the gate voltage so that the
Fermi level lies in the shaded region of the energy dispersion in
Fig.~\ref{Fig.Datta-Das}. 

The Hamiltonian~(\ref{H}) can be written as a matrix in the spin-space
and its eigenvalues are given by $E_{\pm}(k)=\frac{\hbar^2k^2}{2m^*}\pm \sqrt{\D^2 + \al^2k^2}$.
Here we consider two limits, \mbox{$\D\gg\Eso$} and
\mbox{$\Eso\gg\D$}, that allow analytical solutions.


\vspace{1mm}
\emph{Case $\D\gg\Eso$}.
We find it convenient to introduce the dimensionless ratio \mbox{$\gamma=\Eso/\D$}. 
When $\g \ll 1$ the expression for
eigen-energies above
can be simplified as:
\bq
E_\pm(k) \approx \frac{\hbar^2k^2}{2m^*}(1\pm 2\g) \pm \D + \mathcal{O}(\g^2).
\eq
\\[-4mm]
The corresponding eigenfunctions have the form 
\bea
\vert\psi_{-}(k) \ra & 
=&\frac{1}{\sqrt{2(1+\g^2k^2/\kr^2)}}\left(\begin{array} {c}
             1+i\g k/\kr \\
              1-i\g k/\kr
              \end{array}\! \right) e^{i kx} \label{psi_pm} \\
\vert\psi_{+}(k) \ra & 
=&\frac{1}{\sqrt{2(1+\g^2k^2/\kr^2)}}\left(\!  \begin{array} {c}
              1+i\g k/\kr     \\
              -(1-i\g k/\kr)
              \end{array}\! \right) e^{i kx}.
\nonumber
\eea


We aim to obtain the scattering matrix of the system from the
continuity conditions for the wave-functions and their first
derivatives at the boundary between regions I and II, and
II and III of the wire (see Fig.~\ref{Fig.Datta-Das}). The solution in the
Rashba-active region is a linear combination of
$|\psi_-\ra$ and $|\psi_+\ra$, whereas in regions I and III, where
only the Zeeman term is present, it is given by 
spinors $\ups_x$ and $\dns_x$. 
%
In the latter regions the Fermi wave-vectors are
$\kfupdn=k_0\sqrt{1\mp\D/\Ef}$,  
where $k_0=\sqrt{2m^*\Ef}/\hbar$ is the
Fermi wave-vector without magnetic field, and $\D\ll\Ef$ is
assumed throughout.
The corresponding Fermi velocities are given by 
$\vfupdn\!\! =\! v_0\sqrt{1\mp\D/\Ef}$, whereas
for the Rashba-active region II we obtain
$k_{1,2}=\kfupdn/\sqrt{1\pm 2\g}$. 





Resorting to the above expressions for the wave-vectors and velocities
at the Fermi level, one obtains the scattering matrix following 
a standard procedure~\cite{Datta-Das,Egues03}.
The transmission coefficient for the equal-spin polarizations 
in both leads is readily expressed
through the elements of the scattering matrix (neglecting higher
orders in $\g$):
\bea
\Tuu=\Tdd&\approxeq&
1-\frac{4\beta_1\beta_2}{(1+\beta_1\beta_2)^2}\sin^2\left[ \frac{(k_2-k_1)L}{2}
  \right] \label{Tuu1}\\
&\approx& 1-\frac{4\beta_1\beta_2}{(1+\beta_1\beta_2)^2} \sin^2\left[
 \left(\frac{\D}{2\Ef}+\g\right)k_0L \right], \nonumber
\eea
where 
the coefficients $\beta_{1,2}$ are given by
\bq
\beta_{1,2}\equiv\frac{\gamma k_{1,2}}{k_R}\approx
\frac{\sqrt{\Eso\Ef}}{\D}\left(1\mp\frac{\D}{2\Ef} \mp\g\right) .
\label{beta}
\eq

The argument of the sine in Eq.~(\ref{Tuu1}) is the phase shift
between the two subbands, similar to the original Datta-Das
setup. 
However unlike in the former case, the magnetic
field enters Eq.~(\ref{Tuu1}) through the Fermi
wave-vectors $k_{1,2}$, leading to the oscillations of the
transmission  with the field  strength $\Delta$.
It is important to distinguish these oscillations from the well-known
Aharonov--Bohm effect~\cite{AB} and the spin-orbit induced Berry phase~\cite{ALG} 
that stem from the electron wave-function acquiring a phase 
along a closed trajectory in the magnetic field. 


It is evident from Eqs.~(\ref{Tuu1}) and (\ref{beta}) that
in the absence of Rashba effect, $\Eso\to 0$, 
the prefactor in front of the sine vanishes and  $\Tuu\to1$,
as it should be for the fully spin-polarized case.
The physical meaning of Eq.~(\ref{Tuu1}) is thus clear: 
it describes the deviation from the perfect
transmission caused by a finite Rashba interaction. 

Similarly, one can write down the expressions for the cross-channel
transmissions, i.e. for the opposite spin-polarizations in the left 
and right leads~\cite{note1}:
\bea
\Tud&=&\frac{4\beta_1^2}{(1+\beta_1\beta_2)^2}\left(\frac{\vfup}{\vfdn}\right)
\sin^2\left[ \frac{(k_2-k_1)L}{2}  \right]\nonumber\\
\Tdu&=&\frac{4\beta_2^2}{(1+\beta_1\beta_2)^2}\left(\frac{\vfdn}{\vfup}\right)
\sin^2\left[ \frac{(k_2-k_1)L}{2}  \right].
\eea
Here $\Tud\neq\Tdu$ due to the different Fermi velocities and
wave-vectors of the two subbands (see Fig.~\ref{Fig.Datta-Das}).
%

The dependence of $\Tuu$ on the strength of the Rashba
constant $\alpha$ is shown
in Fig.~\ref{Fig.ZeemanR_A} for several values of $\nu\equiv1/\g$,
where we chose $\Ef=0.1$~eV along the lines of Ref.~\cite{Nitta} and the length of the wire $L=400$~nm. $\Tuu(\al)$ is an oscil\-la\-ting function with a striking dependence
of both the amplitude and the period on the magnetic field. 




\bef 
\begin{center}{\includegraphics[width=8.2cm]{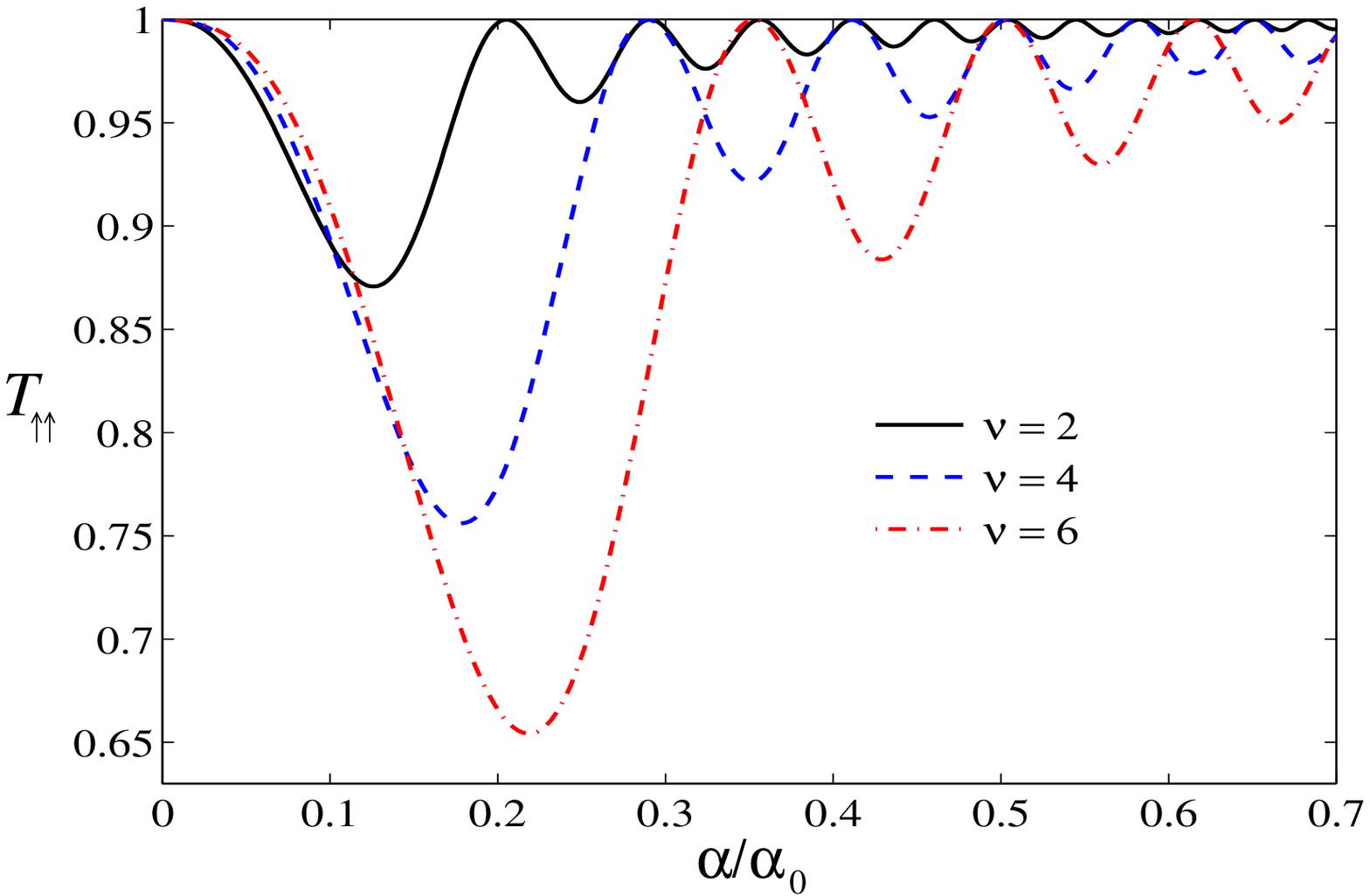}}
\end{center}
\vspace{-7mm}
\caption
{Transmission $\Tuu$ in the equal-spin channel of the Datta-Das device
as a function of the Rashba coefficient~$\alpha$. Three different values of the
 applied magnetic field  \mbox{$\nu\equiv\D/\Eso>1$} have been considered,
 as shown in the legend.
} 
\label{Fig.ZeemanR_A}
\vspace{-5mm}
\ef

Most importantly, instead of varying the strength of the Rashba interaction $\alpha$ by
gating the Datta-Das device, as has been proposed
originally\cite{Datta-Das}, our setup offers another
efficient way of controlling the current spin polarization by means of
varying the applied magnetic field. The dependence of the transmission
$\Tuu$ on the magnetic field $\nu\equiv\D/\Eso$ is shown
with the solid line in Fig.~\ref{Fig.ZeemanR_Z}. 

Let us analyze in more detail the phase shift in Eq.~(\ref{Tuu1}):
\vspace{-3mm}
\bq
\frac{(k_2-k_1)L}{2}\approx k_0L\left(\frac{\D}{2\Ef}+\g\right)\propto(\eso\nu+1/\nu), 
\label{angle2}
\vspace{-2mm}
\eq
where $\eso\equiv\Eso/2\Ef$.
At small values of \mbox{$1\ll\nu<\nu_0$}, with $\nu_0\equiv\sqrt{1/\eso}$,
the second term in
Eq.~(\ref{angle2}) dominates and the transmission is governed by the term
$\sin^2(k_0L/\nu)$, hence the period of oscillations becoming
smaller as $\nu$ decreases (inset in
Fig.~\ref{Fig.ZeemanR_Z}). 
For large values of $\nu$ the linear term in~(\ref{angle2}) prevails, leading to the constant period $1/\eso$
of oscillations. Their amplitude  is
governed by the prefactor of the sine in Eq.~(\ref{Tuu1}), whose
dependence on $\nu$ is shown in broken line in
Fig.~\ref{Fig.ZeemanR_Z}. 

Let us estimate the value of magnetic field $B_R$ when the Zeeman
splitting is equal to the Rashba energy (i.e. $\g=1$). Taking the value of 
$\alpha_0=7\times10^{-10}$~eV~cm, as determined for the InGaAs/InAlAs interface
in Ref.~\cite{Nitta}, 
we obtain the spin-orbit energy
$\Eso=1.6\times10^{-5}$~eV, which, given the electron g-factor $g=4$, corresponds to the magnetic
field $B_R=0.14$~T.
The optimal value of the magnetic field for the situation shown in
Fig.~\ref{Fig.ZeemanR_Z} is $\nu_0\approx40$, corresponding to
the magnetic field $B_0=\nu_0 B_R\approx5.6$~T, which lies in the
experimentally available range. 





\emph{Case $\Eso\gg\D$}.
We now turn our attention to the opposite limit of dominant Rashba interaction and small Zeeman
perturbation ($\nu\equiv\D/\Eso\ll 1)$. We shall construct a
perturbative scheme starting from the $\nu=0$ (pure Rashba)
Hamiltonian in the basis of plane waves:
$H_x=\frac{\hbar^2k^2}{2m^*}-\alpha\hs_y k,\,$
where, as before, we have limited ourselves to only the lowest
transverse subband. For $\nu=0$ the eigenvalues are
$\eps_\pm=\frac{\hbar^2k^2}{2m^*}\mp\alpha k$ and the corresponding 
eigenfunctions $\plus_y$, $\minus_y$ are the eigenstates of
$\hs_y$.

\bef [!tbp]
\begin{center}{\includegraphics[width=8.2cm]{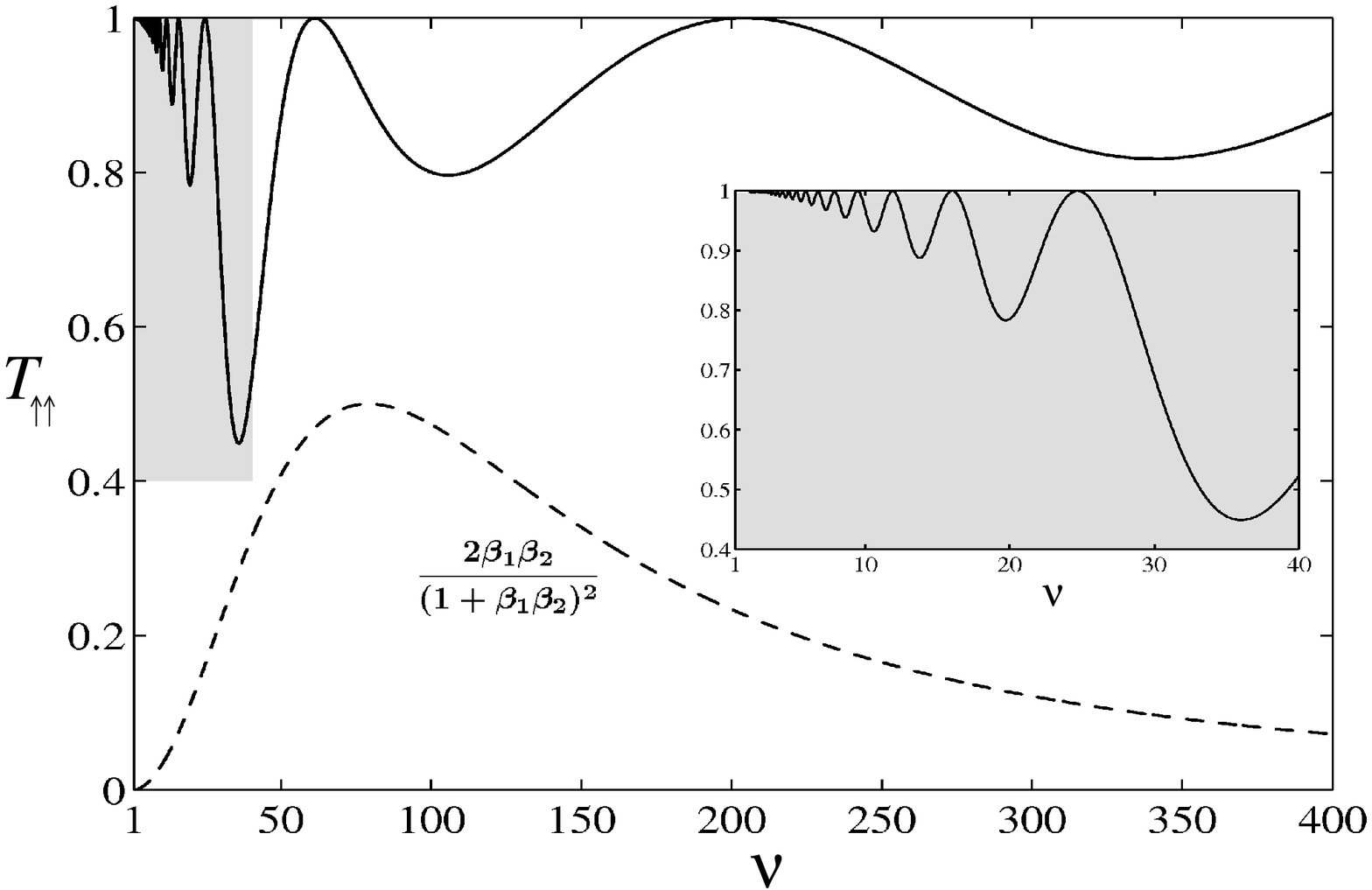}}
\end{center}
\vspace{-7mm}
\caption
{Transmission of the Datta-Das device $T_{\uu}$ (solid line), as a function of magnetic field 
 $\nu=\D/\Eso\gg 1$. The Rashba constant $\alpha=\alpha_0$ was assumed
 as for InGaAs/InAlAs interface in Ref.~\cite{Nitta}. 
 The inset shows the details of the behaviour of $T_{\uu}$ at low
 magnetic fields, in the shaded region of the main plot.
The dashed curve traces the dependence of the numerical prefactor in
 front of $2\sin^2[(k_2-k_1)L/2]$ in $\Tuu$.
}
\label{Fig.ZeemanR_Z}
\vspace{-4mm}
\ef

In terms of these eigenstates, the matrix elements of the Zeeman
perturbation, $V\equiv\D\,\hs_x$, are equal to 
$_y\la-|V|+ \ra_y=-\,_y\la + |V|- \ra_y=i\D$. 
Solving the resulting Hamiltonian at the  2$^\text{nd}$ order of the
perturbation theory, we obtain eigenvalues
$E_\pm\approx \eps_\pm \pm \frac{\D^2}{\eps_+ - \eps_-} 
\equiv \eps_\pm \mp \left(\frac{\varkappa}{k}\right)\D$,
where \mbox{$\varkappa=\frac{\D}{2\alpha}\equiv k_R \nu/4$} is a characteristic
wave-vector with the effective kinetic energy 
$\frac{\hbar^2\varkappa^2}{2m^*} = \nu\Eso/4 \equiv \D/4$.

Of course, the perturbation theory is only valid when the correction
$\delta E=\D\,(\varkappa/k)$ is small compared to bare Rashba energies
$\eps_\pm$, which translates into 
$k\gg\varkappa$. Therefore, in order for the perturbation
theory to be applicable, the Fermi level has to be higher than $ E_{F_\text{min}}\approxeq2\D$.

The eigenstates of the problem at the second
order of the perturbation theory in $\D$ yield the forms 
%
\vspace{-1mm}
\bea
\plus\!\!&=\plus_y - i\frac{\varkappa}{k}\minus_y\!\! &\equiv 
\frac{1}{\sqrt{2(1+\varkappa^2/k^2)}} \left(\!
\ba{c} 1-i\varkappa/k \\ 
       i-\varkappa/k
\ea\! \right)  \label{plus-minus}\\[-1mm]
\minus\!\!&=\minus_y - i\frac{\varkappa}{k}\plus_y\!\! &\equiv \frac{1}{\sqrt{2(1+\varkappa^2/k^2)}}\left(\! 
\ba{c} 1-i\varkappa/k \\ 
       -(i-\varkappa/k)
\ea \right).
\nonumber
\vspace{-2mm}
\eea
%
We note that the bands described by these states are not continuous at
$k=0$, since the perturbation theory is not applicable there.
Consequently, the right-moving
carriers at the Fermi level with $k=k_2$ are described by the
$\plus_{k_2}$ state, whereas the left-moving carriers with the
opposite momentum  are described by the $\minus_{-k_2}$
state. Here the Fermi wave-vectors of the Rashba subbands are given by
\vspace{-3mm}
\bq
%
%
k_{1,2} = k_0\mp k_R + \mathcal{O}\left(\frac{\Eso}{\Ef}, \frac{\D^2}{E_F\Eso} \right)
 \approx k_0(1\mp\beta),
\label{kv_Rashba}
\vspace{-1mm}
\eq
where $\beta\equiv\ k_R/k_0 =\sqrt{2\eso}$ and we neglected 
the higher order terms 
in the last equality.
Using Eqs.~(\ref{plus-minus}--\ref{kv_Rashba}), it is a
straightforward though tedious task 
to write down the boundary
conditions between the Rashba-active and Zeeman-only regions
and obtain the scattering matrix. 
The final result for $\Tuu$ is
given in full in the endnote~\cite{note2}.

The dependence of $\Tuu$ on the Rashba coupling
constant $\alpha$ is illustrated in Fig.~\ref{Fig.RashbaZ}a.
For small magnetic fields $\nu$, the curve is
almost indistinguishable from the cosine (dash-dotted line in
Fig.~\ref{Fig.RashbaZ}a).
%
%
Indeed, one can Taylor-expand $\Tuu(\nu)$ around $\nu=0$ to obtain,
up to linear order,
\bq
\Tuu(\nu)\approx\cos^2\tR - \sin^2\tR\left(
2\cos^2\tR-\eta\sin2\tR\cot\phi \right)\nu, 
\label{Tulin}
\vspace{-1mm}
\eq
where the angle $\phi\equiv k_0L(1+\eso)$ and
$\eta$ is a spin-orbit dependant constant
\mbox{$\eta\equiv\frac{\beta}{2}\frac{1+\eso}{(1+\eso)^2-\beta^2}$}.
Therefore for zero magnetic field we find $\Tuu\big\vert_{\nu=0}=\frac{1}{2}(1+\cos2\tR)$,
in agreement with the conventional Datta-Das formula.

\bef 
\begin{center}{\includegraphics[width=8.6cm]{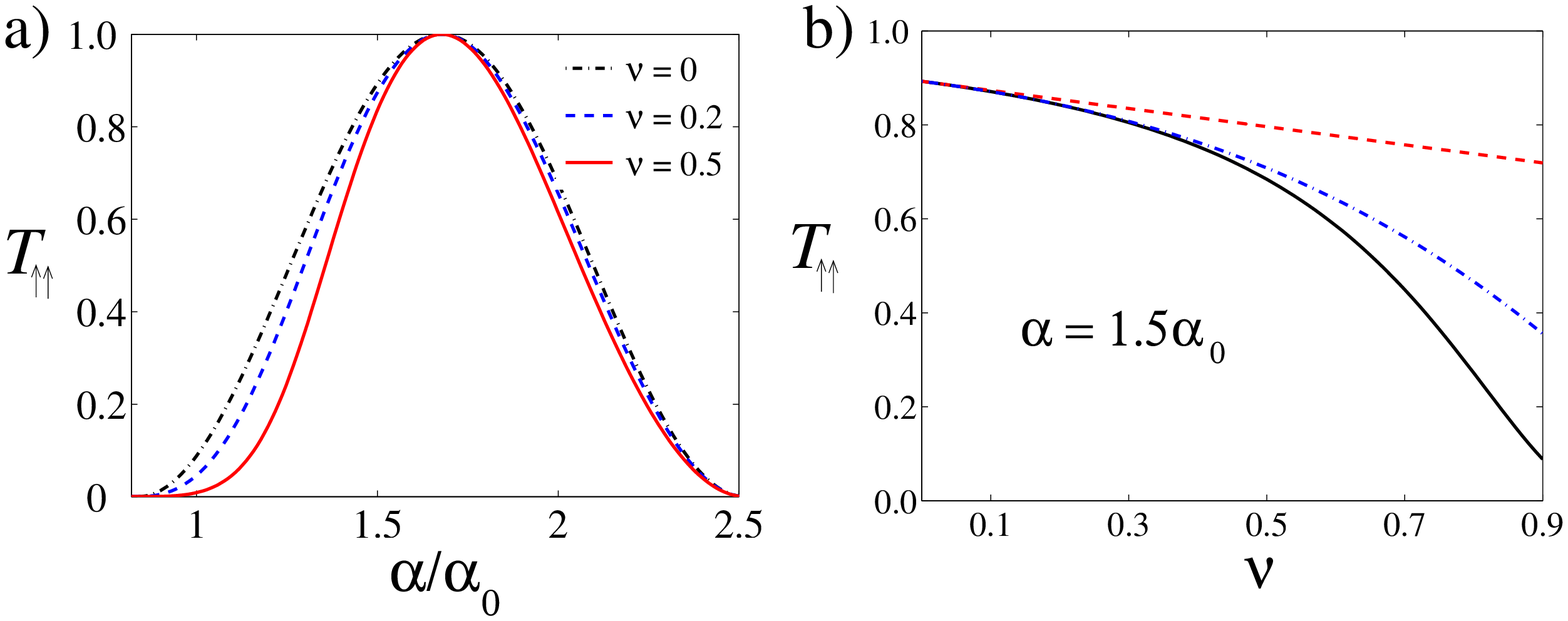}}
\end{center}
\vspace{-7mm}
\caption
{
Transmission of the Datta-Das device,
 $T_{\uu}$, in the limit of weak magnetic field $\nu=\D/\Eso \ll 1$ 
 as a function of (a)
  the Rashba coefficient $\alpha$, for three different values
 of magnetic field (see legend); (b) the 
 magnetic field $\nu$ for a fixed value of $\alpha=1.5\alpha_0$. The
 value of $\alpha_0$ is for the InGaAs/InAlAs interface~\cite{Nitta}.  
 In panel (b), the linear approximation Eq.~(\ref{Tulin}) for
 $\Tuu(\nu)$ is plotted in broken line; the dash-dotted line shows 
 the Taylor expansion of  $\Tuu(\nu)$ up to cubic order in $\nu$. 
} 
\label{Fig.RashbaZ}
\vspace{-4mm}
\ef

The dependence of $\Tuu$ on $\nu$ is plotted in
Fig.~\ref{Fig.RashbaZ}b in solid line.
The linear fit of Eq.~(\ref{Tulin}) is shown in
broken line, and serves as a good approximation for $\nu\lesssim0.2$. 
This regime, $\D\ll\Eso$, turns out
to be less interesting from the application point of view since the
only effect of the magnetic field is to
monotonously suppress the transmission coefficient $\Tuu$ (see Fig.~\ref{Fig.RashbaZ}b), as opposed
to the ubiquitous oscillating behaviour found for large fields.

\emph{Conclusions.}
We have investigated the interplay between the Rashba
spin-orbit interaction and Zeeman effect in quantum wires in the
Datta-Das geometry, exploiting the scattering matrix approach.  
The conventional setup, without an external magnetic field, permits in
principle to control the electron spin polarization by means of variable spin-orbit
interaction $\alpha$. 
However, it has a limited application in practice since the range of
variation of $\alpha$ is quite narrow~\cite{Nitta}.
By adding an in-plane magnetic field to the
setup, we have shown that the prolific combination of the Zeeman
effect with the Rashba spin-orbit interaction offers an unprecedented
control over the electron spin polarization. 
The magnetic field induces characteristic oscillations of the
electron spin polarization, depicted in
Fig.~\ref{Fig.ZeemanR_Z}, with their amplitude
and period depending on the ratio $\nu=\D/\Eso$.
The most promising realm for the operation of
such a spin control device appears to be for Zeeman fields
\mbox{$\Eso\ll\D\ll E_F$}. 
Furthermore, given the typical value of the spin-orbit
interaction reported for the InGaAs/InAlAs interface\cite{Nitta}, we
obtain the optimal value of the magnetic field $B\sim5$~T that is certainly well
accessible experimentally.

The effect of the electron-electron interaction in the Datta-Das setup 
can be potentially interesting~\cite{Devillard05,Gritsev05} and will be a
subject of a future work.
Equally, study of the effect of the higher transversal subbands~\cite{Egues03}
on the conductance in the Zeeman field seems worthwhile.
%

A.H.N. acknowledges the support of FQRNT and
K.L.H. was supported by CIAR, FQRNT, and NSERC.
\vspace{-3mm}


\begin{thebibliography}{22}
%
\expandafter\ifx\csname natexlab\endcsname\relax\def\natexlab#1{#1}\fi
\expandafter\ifx\csname bibnamefont\endcsname\relax
  \def\bibnamefont#1{#1}\fi
\expandafter\ifx\csname bibfnamefont\endcsname\relax
  \def\bibfnamefont#1{#1}\fi
\expandafter\ifx\csname citenamefont\endcsname\relax
  \def\citenamefont#1{#1}\fi
\expandafter\ifx\csname url\endcsname\relax
  \def\url#1{\texttt{#1}}\fi
\expandafter\ifx\csname urlprefix\endcsname\relax\def\urlprefix{URL }\fi
\providecommand{\bibinfo}[2]{#2}
\providecommand{\eprint}[2][]{\url{#2}}

\bibitem[{\citenamefont{Awschalom et~al.}(2002)\citenamefont{Awschalom, Loss,
  and Samarth}}]{spintronics}
\bibinfo{editor}{\bibfnamefont{D.~D.} \bibnamefont{Awschalom}},
  \bibinfo{editor}{\bibfnamefont{D.}~\bibnamefont{Loss}}, \bibnamefont{and}
  \bibinfo{editor}{\bibfnamefont{N.}~\bibnamefont{Samarth}}, eds.,
  \emph{\bibinfo{title}{Semiconductor spintronics and quantum computation}}
  (\bibinfo{publisher}{Springer, Berlin}, \bibinfo{year}{2002}).

\bibitem[{\citenamefont{Bennet and DiVicenzo}(2000)}]{qc_review00}
\bibinfo{author}{\bibfnamefont{C.~H.} \bibnamefont{Bennet}} \bibnamefont{and}
  \bibinfo{author}{\bibfnamefont{D.~P.} \bibnamefont{DiVicenzo}},
  \bibinfo{journal}{Nature (London)} \textbf{\bibinfo{volume}{404}},
  \bibinfo{pages}{247} (\bibinfo{year}{2000}).

\bibitem[{\citenamefont{Datta and Das}(1990)}]{Datta-Das}
\bibinfo{author}{\bibfnamefont{S.}~\bibnamefont{Datta}} \bibnamefont{and}
  \bibinfo{author}{\bibfnamefont{B.}~\bibnamefont{Das}},
  \bibinfo{journal}{Appl. Phys. Lett.} \textbf{\bibinfo{volume}{56}},
  \bibinfo{pages}{665} (\bibinfo{year}{1990}).

\bibitem[{\citenamefont{Bychkov and Rashba}(1984)}]{Rashba}
\bibinfo{author}{\bibfnamefont{Y.~A.} \bibnamefont{Bychkov}} \bibnamefont{and}
  \bibinfo{author}{\bibfnamefont{E.~I.} \bibnamefont{Rashba}},
  \bibinfo{journal}{JETP Lett.} \textbf{\bibinfo{volume}{39}},
  \bibinfo{pages}{78} (\bibinfo{year}{1984}).

\bibitem[{\citenamefont{Egues et~al.}(2002)\citenamefont{Egues, Burkard, and
  Loss}}]{Egues02}
\bibinfo{author}{\bibfnamefont{J.~C.} \bibnamefont{Egues}},
  \bibinfo{author}{\bibfnamefont{G.}~\bibnamefont{Burkard}}, \bibnamefont{and}
  \bibinfo{author}{\bibfnamefont{D.}~\bibnamefont{Loss}},
  \bibinfo{journal}{Phys. Rev. Lett.} \textbf{\bibinfo{volume}{89}},
  \bibinfo{pages}{176401} (\bibinfo{year}{2002}).


\bibitem[{\citenamefont{Serra et~al.}(2005)\citenamefont{Serra, S\'anchez, and
  L\'opez}}]{Serra05}
\bibinfo{author}{\bibfnamefont{L.}~\bibnamefont{Serra}},
  \bibinfo{author}{\bibfnamefont{D.}~\bibnamefont{S\'anchez}},
  \bibnamefont{and} \bibinfo{author}{\bibfnamefont{R.}~\bibnamefont{L\'opez}},
  \bibinfo{journal}{Phys. Rev. B} \textbf{\bibinfo{volume}{72}},
  \bibinfo{pages}{235309} (\bibinfo{year}{2005}).


\bibitem[{\citenamefont{Devillard et~al.}(2005)\citenamefont{Devillard,
  Cr\'epieux, Imura, and Martin}}]{Devillard05}
\bibinfo{author}{\bibfnamefont{P.}~\bibnamefont{Devillard}},
  \bibinfo{author}{\bibfnamefont{A.}~\bibnamefont{Cr\'epieux}},
  \bibinfo{author}{\bibfnamefont{K.~I.} \bibnamefont{Imura}}, \bibnamefont{and}
  \bibinfo{author}{\bibfnamefont{T.}~\bibnamefont{Martin}},
  \bibinfo{journal}{Phys. Rev. B} \textbf{\bibinfo{volume}{72}},
  \bibinfo{pages}{041309(R)} (\bibinfo{year}{2005}); 
  %
  \bibinfo{author}{\bibfnamefont{H.}~\bibnamefont{Lee}} \bibnamefont{and}
  \bibinfo{author}{\bibfnamefont{S.-R.} \bibnamefont{Yang}},
  \bibinfo{journal}{\emph{ibid.}} \textbf{\bibinfo{volume}{72}},
  \bibinfo{pages}{245338} (\bibinfo{year}{2005}).



\bibitem{Cahay03}
\bibinfo{author}{M. Cahay and S. Bandyopadhyay},
\bibinfo{journal}{Phys. Rev. B} \textbf{\bibinfo{volume}{68}},
  \bibinfo{pages}{115316} (\bibinfo{year}{2003}); \emph{ibid.}
  \textbf{\bibinfo{volume}{69}}, \bibinfo{pages}{045303} (2004).


\bibitem{AB}
\bibinfo{author}{Y. Aharonov and D. Bohm},
\bibinfo{journal}{Phys. Rev.} \textbf{\bibinfo{volume}{115}},
  \bibinfo{pages}{485} (\bibinfo{year}{1959}).

\bibitem{ALG}
\bibinfo{author}{A. G. Aronov and Y. B. Lyanda-Geller},
\bibinfo{journal}{Phys. Rev. Lett.} \textbf{\bibinfo{volume}{70}},
  \bibinfo{pages}{343} (\bibinfo{year}{1993}).

\bibitem[{\citenamefont{Landau and Lifshitz}(1991)}]{Landau}
\bibinfo{author}{\bibfnamefont{L.~D.} \bibnamefont{Landau}} \bibnamefont{and}
  \bibinfo{author}{\bibfnamefont{E.~M.} \bibnamefont{Lifshitz}},
  \emph{\bibinfo{title}{Quantum mechanics}} (\bibinfo{publisher}{Pergamon
  Press, Oxford}, \bibinfo{year}{1991}).

\bibitem[{\citenamefont{Das et~al.}(1990)\citenamefont{Das, Datta, and
  Reifenberger}}]{Das90}
\bibinfo{author}{\bibfnamefont{B.}~\bibnamefont{Das}},
  \bibinfo{author}{\bibfnamefont{S.}~\bibnamefont{Datta}}, \bibnamefont{and}
  \bibinfo{author}{\bibfnamefont{R.}~\bibnamefont{Reifenberger}},
  \bibinfo{journal}{Phys. Rev. B} \textbf{\bibinfo{volume}{41}},
  \bibinfo{pages}{8278} (\bibinfo{year}{1990}).

\bibitem[{\citenamefont{Chen et~al.}(1993)\citenamefont{Chen, Han, Huang,
  Datta, and Janes}}]{Chen93}
\bibinfo{author}{\bibfnamefont{G.~L. Chen} \emph{et~al.}},
  \bibinfo{journal}{Phys. Rev. B} \textbf{\bibinfo{volume}{47}},
  \bibinfo{pages}{4084} (\bibinfo{year}{1993}).

\bibitem[{\citenamefont{Dresselhaus}(1955)}]{Dresselhaus}
\bibinfo{author}{\bibfnamefont{G.}~\bibnamefont{Dresselhaus}},
  \bibinfo{journal}{Phys. Rev.} \textbf{\bibinfo{volume}{100}},
  \bibinfo{pages}{580} (\bibinfo{year}{1955}).

\bibitem[{\citenamefont{Luo et~al.}(1990)\citenamefont{Luo, Munekata, Fang, and
  Stiles}}]{Luo90}
\bibinfo{author}{\bibfnamefont{J.}~\bibnamefont{Luo}},
  \bibinfo{author}{\bibfnamefont{H.}~\bibnamefont{Munekata}},
  \bibinfo{author}{\bibfnamefont{F.~F.} \bibnamefont{Fang}}, \bibnamefont{and}
  \bibinfo{author}{\bibfnamefont{P.~J.} \bibnamefont{Stiles}},
  \bibinfo{journal}{Phys. Rev. B} \textbf{\bibinfo{volume}{41}},
  \bibinfo{pages}{7685} (\bibinfo{year}{1990}).

\bibitem[{\citenamefont{Nitta et~al.}(1997)\citenamefont{Nitta, Akazaki, and
  Takayanagi}}]{Nitta}
\bibinfo{author}{\bibfnamefont{J.}~\bibnamefont{Nitta}},
  \bibinfo{author}{\bibfnamefont{T.}~\bibnamefont{Akazaki}}, 
  \bibinfo{author}{\bibfnamefont{H.}~\bibnamefont{Takayanagi}}, \bibnamefont{and}
  \bibinfo{author}{\bibfnamefont{T.}~\bibnamefont{Enoki}}
  \bibinfo{journal}{Phys. Rev. Lett.} \textbf{\bibinfo{volume}{78}},
  \bibinfo{pages}{1335} (\bibinfo{year}{1997}).

\bibitem[{\citenamefont{Hassenkam et~al.}(1997)\citenamefont{Hassenkam,
  Pedersen, Baklanov, Kristensen, Sorensen, Lindelof, Pikus, and
  Pikus}}]{Hassenkam97}
\bibinfo{author}{\bibfnamefont{T. Hassenkam} \emph{et~al.}},
  \bibinfo{journal}{Phys. Rev. B} \textbf{\bibinfo{volume}{55}},
  \bibinfo{pages}{9298} (\bibinfo{year}{1997}).



\bibitem[{\citenamefont{Moroz and Barnes}(1999)}]{Moroz99}
\bibinfo{author}{\bibfnamefont{A.~V.} \bibnamefont{Moroz}} \bibnamefont{and}
  \bibinfo{author}{\bibfnamefont{C.~H.~W.} \bibnamefont{Barnes}},
  \bibinfo{journal}{Phys. Rev. B.} \textbf{\bibinfo{volume}{60}},
  \bibinfo{pages}{14272} (\bibinfo{year}{1999}).

\bibitem[{\citenamefont{Governale and Zulicke}(2004)}]{Governale04}
\bibinfo{author}{\bibfnamefont{M.}~\bibnamefont{Governale}} \bibnamefont{and}
  \bibinfo{author}{\bibfnamefont{U.}~\bibnamefont{Zulicke}},
  \bibinfo{journal}{Solid State Communications} \textbf{\bibinfo{volume}{131}},
  \bibinfo{pages}{581} (\bibinfo{year}{2004}).

\bibitem[{\citenamefont{Egues et~al.}(2003)\citenamefont{Egues, Burkard, and
  Loss}}]{Egues03}
\bibinfo{author}{\bibfnamefont{J.~C.} \bibnamefont{Egues}},
  \bibinfo{author}{\bibfnamefont{G.}~\bibnamefont{Burkard}}, \bibnamefont{and}
  \bibinfo{author}{\bibfnamefont{D.}~\bibnamefont{Loss}},
  \bibinfo{journal}{App. Phys. Lett.} \textbf{\bibinfo{volume}{82}},
  \bibinfo{pages}{2658} (\bibinfo{year}{2003}).

\bibitem{note1}
The spins in the
leads are assumed to be collinear, for the treatment of non-collinear lead magnetizations
see e.g., M. H. Larsen, A. M. Lunde, and K. Flensberg, Phys. Rev.~B {\bf66},
033304 (2002).

\bibitem{note2}
%
For the case  $\Eso\gg\D$, the transmission in the equal-spin channel
is
%
\vspace{-3mm}
\protect\begin{displaymath}
\!\!\!\Tuu=\frac{4 (1-\nu)^2
  F(\theta_R)^2}{c^2(1-\nu)^2-2c(1-\nu)P(\theta_R,\nu)+|Q(\theta_R,\nu)|^2},\!\!
\label{Tuu2} 
\vspace{-1mm}
\protect\end{displaymath}
where the auxiliary functions $F$, $P$ and $Q$
depend on the Rashba angle $\tR\equiv k_RL$ and angle
$\phi\equiv k_0L(1+\eso)$ as\\[1mm]
%
%
$
\protect\ba{lcl}
F(\tR)&=&b \cos\phi \sin\tR + c\sin\phi \cos\tR \\
P(\tR,\nu)&=&a \cos2\phi + (b^2+c^2\nu)\cos2\phi \cos2\tR\\
& & - \,b\, c\,(1-\nu) \sin2\phi\sin2\tR \\
Q(\tR,\nu)&=& a + \frac{1}{2}\left[e^{i2\phi}(b+c)(b-c\nu)+{}\right.\\
& {}&+ \left. e^{-i2\phi}(b-c)(b+c\nu) \right],
\protect\ea
$
\vspace{1mm}\\
%
where the following notations have been used:\\[1mm]
$a=(1-\g_1^2)(1-\g_2^2), \; b=\g_1+\g_2, \; c=1+\g_1\g_2$  \\
$\g_{1,2}=\frac{\varkappa}{k_{1,2}}=\frac{\nu}{4}\frac{k_R}{k_0(1+\eso\mp\beta)}
\equiv\frac{\nu\beta}{4(1+\eso\mp\beta)}$.
%


\bibitem[{\citenamefont{Gritsev et~al.}(2005)\citenamefont{Gritsev, Japaridze,
  Pletyukhov, and Baeriswyl}}]{Gritsev05}
\bibinfo{author}{\bibfnamefont{V.}~\bibnamefont{Gritsev}},
  \bibinfo{author}{\bibfnamefont{G.}~\bibnamefont{Japaridze}},
  \bibinfo{author}{\bibfnamefont{M.}~\bibnamefont{Pletyukhov}},
  \bibnamefont{and}
  \bibinfo{author}{\bibfnamefont{D.}~\bibnamefont{Baeriswyl}},
  \bibinfo{journal}{Phys. Rev. Lett.} \textbf{\bibinfo{volume}{94}},
  \bibinfo{pages}{137207} (\bibinfo{year}{2005}).

\end{thebibliography}

\end{document}